# MC-curves and aesthetic measurements for pseudospiral curve segments

*Rushan Ziatdinov*
*Department of Computer & Instructional Technologies,*
*Fatih University,34500 Büyükçekmece, Istanbul, Turkey*
*E-mail: rushanziatdinov@gmail.com, ziatdinov@fatih.edu.tr*
*URL: http://www.ziatdinov-lab.com/*

*Rifkat I. Nabiyev*
*Department of Fine Art and Costume Art, Faculty of Design and National Culture,*
*Ufa State University of Economics and Service, 450068 Ufa, Russia*
*E-mail: dizain55@yandex.ru*
*URL: http://www.nabiyev.mathdesign.ru/*

*Kenjiro T. Miura*
*Department of Information Science and Technology,*
*Graduate School of Science and Technology, Shizuoka University,*
*3-5-1, Johoku, Naka-ku, Hamamatsu Shizuoka, 432 Japan*
*E-mail: tmkmiur@ipc.shizuoka.ac.jp*
*URL: http://ktm11.eng.shizuoka.ac.jp/profile/ktmiura/welcome.html*

**Abstract.** This article studies families of curves with monotonic curvature function (MC-curves) and their applications in geometric modelling and aesthetic design. Aesthetic analysis and assessment of the structure and plastic qualities of pseudospirals, which are curves with monotonic curvature function, are conducted for the first time in the field of geometric modelling from the position of technical aesthetics laws. The example of car body surface modelling with the use of aesthetics splines is given.

1. **Introduction**

Aesthetic appeal of industrial products is a very important factor for their successful promotion in the market. Most of the curves and surface profiles used within traditional CAD/CAM systems [1] have polynomial or rational parametric form and do not meet high aesthetic requirements [2]. One of their disadvantages is the difficulty of controlling the monotonicity of the curvature function.

2. **Curves with monotonic curvature function**

Computer-aided geometric design refers to monotone-curvature curves as *fair curves* [3]. Unfortunately, this attitude is based only on well-known geometric principles, and the laws of technical aesthetics are not taken into account. Therefore, in this work we prefer to avoid this term, using "*MC-curve*" (monotone-curvature curve) instead.

MC-curves include spirals with monotonic curvature function (Euler spiral, Nielsen's spiral, logarithmic spiral, involutes of a circle), *pseudospirals* [4] and so-called *log-aesthetic curves* [2], actually a linear reparameterization of pseudospirals. The curves comprise the

*superspiral* [5] family, curvature function of which is given by the Gaussian hypergeometric function satisfying conditions of strict monotonicity with several limitations applied to its parameters [6]. Recently, class A Bézier curves with monotonic curvature function were proposed [7]; detailed analysis has shown, however, that when the polynomial degree is increased the curve restricts to a logarithmic spiral [8]. Controlling the Bézier and B-spline curve's (for polynomial degree *n>2*) curvature function's monotonicity requires deeper analysis and the development of corresponding algorithms. Curves satisfying the monotonicity of curvature function are widely used in car surface design, bevel design, development of transition curves in highways and railroad design, and font design [9-10], and are a crucial element of *aesthetic design* [11]. Figure 1 shows a new conceptual design of a car created with aesthetic curves and *multispirals* [10].

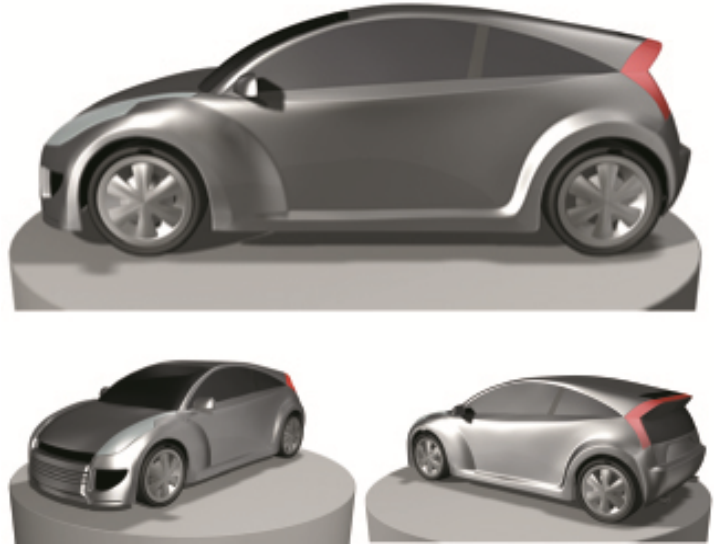

Fig.1. New conceptual design[1] of a car created by log-aesthetic curves, multispirals and kinematic spiral surfaces.

### 3. Mathematical models of log-aesthetic curves

From the perspective of aesthetic design and potential applications, the log-aesthetic curves[2] are the most interesting; their mathematical theory has been elaborated in a number of studies [12-15] [2] [9]. By studying properties of multiple attractively shaped curves' characteristics in real and virtual-world objects, Toshinobu Harada and his research team discovered that their curvature functions were linear or near-linear, depending on the natural parameter (arc length) on the logarithmic curvature graph (LCG) [12-13]. Natural equations of these curves are as follows [8]:

$$\kappa(s) = \begin{cases} e^{-\lambda s}, & \alpha = 0 \\ (\lambda\alpha s + 1)^{-\frac{1}{\alpha}}, & \alpha \neq 0 \end{cases}, \quad (1)$$

where $\alpha$ is the shape parameter and $\lambda$ is the scaling factor. The Gauss-Kronrod numerical integration method has been used for curve segment computation in [2]. In [9], parametric equations for (1) in terms of incomplete gamma functions were found; the obtained equations allowed the computation of curve segments with a high degree of accuracy. A computation experiment staged in [2] yielded drawable regions for the control point, which is determined by the directions of unit tangent vectors in the initial and end points of the aesthetic curve segment. It was found that the Euler (α = -1) and Nielsen (α = 0) spirals, as well as the involutes of a circle (α = 2), have limited drawable regions for the control point, determined by the directions of unit tangent vectors in the initial and end points; the curve segment which fits with tangent directions therefore does not always exist. Nevertheless, judging by the computation experiment carried out in [2], the two-point Hermite interpolation problem always has a solution for spirals with the shape parameter $0 \leq \alpha \leq 1$, but this hypothesis has not been analytically proven. Yoshida and Saito [2] studied the behaviour of reflection lines on ruled spiral surfaces and came to the conclusion that when kinematic surfaces are rotated they lack reflection line oscillations, which is one of the features of their high quality.

---

[1] Created by Prof. Takashi Hada and Tomonobu Nishikawa (Faculty of Design, Shizuoka University of Art and Culture, Japan).

[2] The term *log-aesthetic curve (LAC) is* used in foreign literature as suggested by Prof. Carlo H. Sequin from the University of California (USA).

Some examples of application of $C^1$ class aesthetic splines can be found in [16].

### 4. Unit Quaternion Integral Curves

The class of a *Unit Quaternion Curves* q(s) in the SO(3) rotation group was first introduced in [17]. This study also proposed a method enabling transformation of a curve, determined by a sum of basis functions, into its unit quaternion analogue in SO(3). Given a spline curve in $R^3$, the spline curve is reformulated in a cumulative basis form and the corresponding quaternion curve is constructed by converting each vector addition into the quaternion multiplication. For example, using the method proposed in [17] for splines such as Bézier, Hermite and B-splines, unit quaternion curves can be derived, many differential properties of which are invariant.

*Unit quaternion integral curve (QI-curve)* is defined as:

$$C(s) = P_0 + \int_0^s q(s)\hat{v}_0 q^{-1}(s)ds, \qquad (2)$$

where *s* is the arc length and $\hat{v}_0$ is the arbitrary unit vector [18]. In this case, the quaternions, especially unit quaternions, are useful to describe rotations and are used to control the direction of the tangent vector, which adds some efficiency and simplicity in the design of aesthetic curve shapes. Quaternion coordinates are considered ideal for orientation interpolation of objects [19].

Due to the fact that the QI-curve is defined by a tangent vector, which is controlled by unit quaternion curve, the arc length being its natural parameter, a more convenient manipulation of its curvature function is possible than using the polynomial parametric curves in CAD/CAM systems.

### 5. Historical, theoretical and methodological foundations of science and art integration in the context of aesthetic analysis of pseudospiral curve segments

Since ancient times, the ability to cause sensual empathy and a positive state of mind, and to encourage creative activities, was a matter of learning not only in art, but also many scientific disciplines. History suggests that the search for ways to improve living space has been inspired by knowledge of the world and born of positive spiritual incentives; in broad terms, enjoyment of life is conditional upon the integration of precise scientific disciplines with forms of arts.

Note that this is not to be understood narrowly. These are powerful mechanisms that have formed a certain style of human thinking and modelled human perception of the world, connected to a large extent with the interests of developing certain personality types: ascetic, dogmatic, scholastic, humanistic, etc. Knowledge of the mechanisms influencing the psyche of an individual through art allows the simulation of consciousness, as well as the content approved in the attributes of form through a specific attitude expressed in the formal signs of the results of aesthetic understanding of the world. There is an example from history: "In Ancient Egypt, where one of the means of expressing ruling class ideology was architecture, the chief architect, in addition to his duties, carried out the duties of vizier, keeper of the seal or high priest" [24]. The great monuments of any culture are the first examples illustrating the means of influence on human consciousness, and thus on human mentality. United by their components, these monuments were intended to assert the power of the rulers and establish the norms of peoples' spiritual life.

Thus, art in all cultural and historical periods has had a clearly specified nature, since the psychological potential of its instruments was a strategic mechanism for forming moral and ethical behavioural standards, which determined the content of human spiritual ideals.

It is no coincidence that information could be brought to the human consciousness most effectively if form gave rise to interest and stimulated the perception of its characteristics. That is, this form had to suggest a specific idea to the perceiver as quickly as possible. It is possible to arouse an immediate effect of interest using human psychological mechanisms. The form must have "useful" qualities which can help to actualize the possibilities of effective influence on the psycho-emotional sphere of an individual.

This is about the beauty of form as an expression of the highest indicator of the unity and integrity of all of its components.

In their own way, beauty or disharmony of form in human consciousness causes an instinctive reflex. By comparison with the ethical sphere of personality, we can discover an interrelation in the nature of this reaction: the human reaction to moral aspects can be positive or negative. Such categories as "measure", "beauty" and "harmony", used in evaluating the external quality of form, are also relevant in evaluating an individual's moral qualities.

Methodologically it is important to note that consideration of the question of evaluating the beauty of form was historically based on the union of Art and Science, allowing analyses to be made using two mutually complementary perspectives. This dialectic of interaction between two forms of social consciousness makes it possible to exclude unreasonable declarative conclusions and solve problems in well-reasoned and demonstrative forms.

In view of the above, we emphasize that the best minds of Ancient Greece accepted the idea of achieving harmony in Art by means of exact sciences. Thus, "under the influence of Pythagorean philosophy, which proclaimed 'that all things are regulated and can be learned by the strength of numbers', the Canon of Polykleitos arose in sculpture (5$^{th}$ century BC), and the Grid Plan developed in urban planning". Socrates was the first to formulate the thesis that the beautiful is not just the sum of beautiful parts: "conformity and subordination are required to create artistic unity" [24]. Plato proved the importance of harmonizing the proportions of form, influenced by the discovery of irrationality ($\sqrt{2}$) in Ancient Greek mathematics. It is worth noting that this discovery negated the importance of the mathematical atomism of the Pythagorean School, as its original idea was based on idealistic philosophy.

In this regard, for example, Abu Nasr Al-Farabi, one of the greatest Eastern scientists, while acknowledging the Pythagorean theory of music, rejected the teaching of the Pythagorean connection between music and the movement of the stars as nothing more than a baseless invention. Ibn Sina said about this that he did not seek to establish a link between the state of the sky, the behaviour of the soul, and musical intervals, for "it is the custom of those who cannot distinguish one science from the other" [24].

We should note, however, that, in its purest form, proportioning based on the golden section is rare in the practice of fine art, design, and architecture. Of greater importance for the achievement of aesthetic value and reasonability is to measure the consistency of the "golden section" with other means of expression.

Also of methodological significance in the context of the integration of science and art is the work of the famous Roman architect Vitruvius, who came to identify certain numerical relationships in proportion to the human body, thus creating a scientific basis for ergonomic systems of proportioning. These later gained acceptance in the fine arts and architecture under the name of "The Vitruvian Man". The Italian Renaissance provides an example of cultivating ideas for the unity of science and art in order to nurture a new type of socio-cultural identity, an ethical relationship to the world that is determined by the depth and breadth of true knowledge. It was no accident that such a cultural setting for the disclosure and realization of human artistic potential found itself logically embodied in the outstanding people of the Renaissance, and above all in the work of Leonardo da Vinci, who is known as the pioneer of design.

Thus, in connection with the above, it is possible to come to a definitive conclusion that the union of science and art creates the conditions for nurturing a humanistic style of thought in man which affirms the unity of beauty and usefulness in the spiritual and material space of life, as a leading ethical principle.

### 6. Aesthetic analysis of pseudospiral curve segments

Let us consider the features of the curves (1) presented in the figures, and endeavour to describe their aesthetic properties from the point of view of technical aesthetics laws.

Aesthetic judgement is a way to establish the aesthetic value of an object, to be aware of the result of aesthetic perception, usually fixed in judgements like "It's beautiful", "It's ugly", and so on [25]. In this sense, addressing the evaluation of curves' aesthetic properties should be done in the framework of human nature in terms of aesthetic perception, where latent semantic depth of form can be revealed.

When performing aesthetic evaluation, an important indicator of curves' beauty is harmony. This covers both substantive and formal features of the object and is rated as the highest form of its organization, order, and structural integrity.

It is crucial to detect the mechanisms of making a form integral and harmonious in the totality of its qualities, free from a gustative approach and an idealistic interpretation of beauty. Here we should define the extent of how a person's subjective aesthetic mode of evaluating the formal indices is related to its objective aesthetic value, taking into consideration the relativity aspect in the estimation of an object value. It presupposes that in the framework of one system its aesthetic characteristics might constitute value, while in the framework of the other the value is constituted by its usefulness. The subjective criteria of aesthetic evaluation should be formed on the basis of objective beauty factors. Such a principle will to a huge degree define the objectivity of the judgements made by a person, acting as a recipient in various spacial, material and cultural circumstances of interaction with other objects.

It should be noted that by the concept of "form" we understand the unity of the external and intrinsic properties of an object. In an aesthetic assessment of those properties of form which result in a corresponding human emotional response, one has to identify the objective laws of the design process which are intended to provide integrated and harmonious contours. At the same time one must take into account that, in the design of an object, the form must also comply with the requirements of aesthetic, functional and economic value. In this regard a design item should contain no unjustifiable random elements. This is particularly relevant with decoration which may conceal underlying discordancy and create the effect of false beauty. Such compensations for imperfect form result in a loss of self-sufficiency – an impairment of structure. Here only substantial unity of the decorative attributes as part of the shape itself can provide the aesthetic, ergonomic, economic and social effects which ultimately result in positive value-judgements of the item.

Assessment of the appropriateness of the aesthetic features of curves should be based on the actual properties expressed in the plastic characteristics of those curves. It is noteworthy that applying certain theoretical design elements to them allows the identification of the propriety of their use (Figs. 5, 6).

With regard to the assessment of the aesthetic properties of form, the main issue is to identify those mechanisms in shaping a design object which provide a structural balance of composition and integrity related to perceptions of the attributes of form, since any violation of such structural relationships leads to a deformation of the entire compositional structure.

Since human perception involves visual judgement, compositional laws are perceived by an individual as objectively active conditions for the comfortable perception of an object, taking all its properties as a whole. The objectivity of those conditions is realized through the psychological mechanisms of human perception and their corresponding reactions to specific stimuli: visual, tactile, auditory, etc.

These mechanisms are considered within the framework of perception regularities inherent in all persons due to the common functioning principles of the higher nervous system. Thus, it is known that the process of human reaction within visual perception is transformed into visual judgement "which is not the result of intellectual activity since the latter takes place after the process of perception has been completed" [23]. It logically follows that "the characteristics of objects perceptible to the eye constitute an essential, intrinsic part of the visual process as such" [23].

One such characteristic is "intensity" as a quality of the object's formal properties and a characteristic of the dynamic interaction of internal "forces," one of these being "attraction." Tension, in turn, "has its value and direction – the criteria which can be treated as a psychological force" [23]. The value of intensity depends on the nature of the combination of active "forces", represented as the interaction of the object's attributes resulting in certain qualitative changes between the elements of the composition.

At the same time, the arrangement itself is defined by guides or axes that can be any curves along which the elements are structured and grouped with each other. These theoretical guides may be characterized as representing significant factors for harmonization of relations between composite elements, as well as prerequisites to optimize the properties for formal features of facilities.

Observe the two graphs with curves (1) in Fig. 2. The principle for "structural interaction of curves with coordinate axes" is used as the basis for analysis of curves' aesthetic properties. The curve on the left graph is more homogeneous and stable than the one on the right graph. The laconism of the curve is stipulated by a more visible connection with the coordinate axis: the distance of the upper top is smaller in comparison with the curve on the right example; the attractive force to the axis is more strongly expressed. The feeling of larger stability and integrity is evidence of reasonability and utility, which are transformed into aesthetic reasonability at the level of sensory perception and are characterized by the estimations "more beautiful" and "more pleasant". The low curves in Fig. 2 demonstrate the maximum stress ranges, marked with arrows.

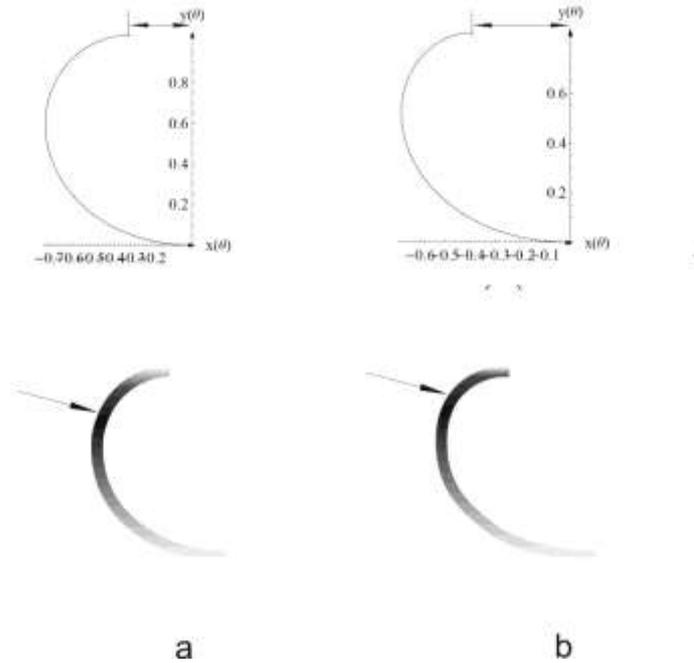

Fig. 2. Comparative aesthetic measurements for Euler spiral ($\alpha = 1$) (a) and Nielsen spiral ($\alpha = 0$) (b).

Fig. 3 shows three plots with examples illustrating the stress range of each curve. Comparing the charts, it is possible to conclude that middle curve (b) is the most compliant with terms of feasibility. It is almost a hemisphere (characterized by a symmetrical shape); hence, it is concise and self-sufficient.

Options (a) and (c) lack the integrity inherent to the curve in example (b). Development of these curves is not finished, with the logical support of another object (required to compensate for their instability with respect to coordinate axes).

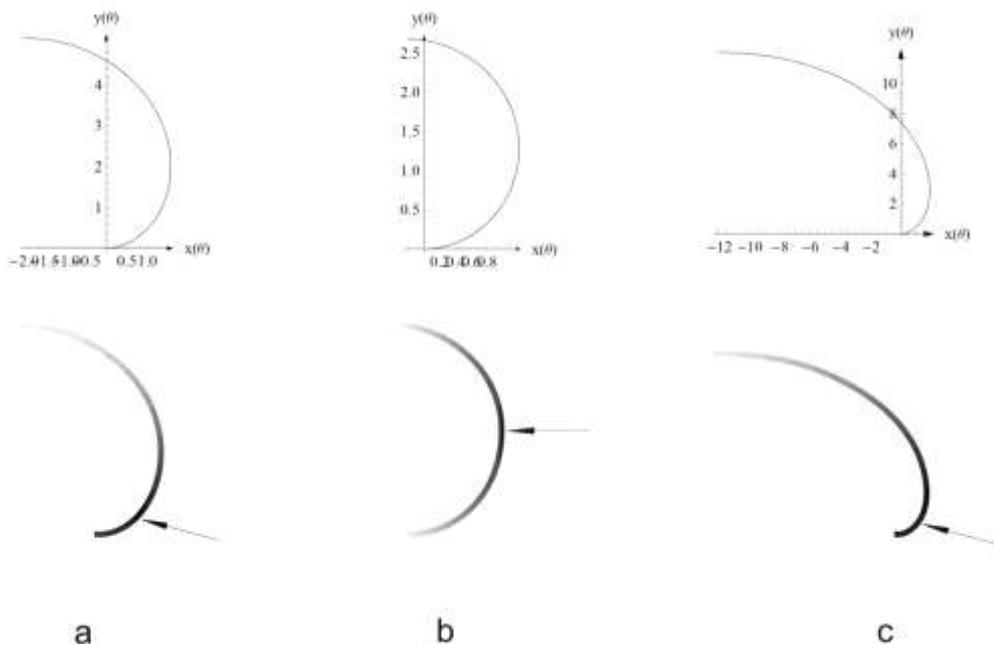

Fig. 3. Comparative aesthetic analysis of plastic properties of the curves: involutes of a circle ($\alpha = 2$) (a) quasi-circle ($\alpha = 10$) (b) and logarithmic spiral ($\alpha = 1$) (c).

Fig. 4 shows an object perception pattern (depending on change of its properties). The figure shows a curve (g) (Euler (Cornu) spiral, ($\alpha = -1$)) with different options of thickness. Upon increase of the line thickness, the visual weight of the curve varies (thus affecting the plastic features). Fig. 4 (e) shows a critical thickness, which does not interfere with feasibility of the curve plastics. Increase in thickness of the curve results in visual deformation of its plastics: the curve is visually compressed. Reduction in thickness of the curve results in loss of structural integrity.

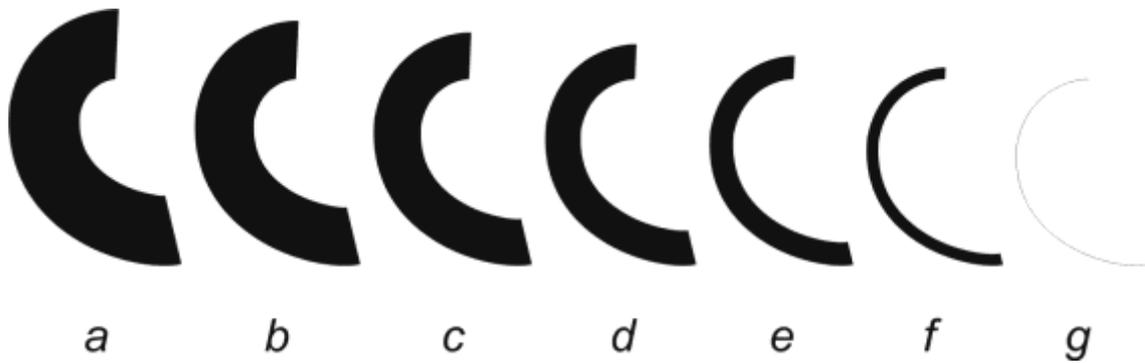

Fig. 4. Object perception depending on change of its features.

Figure 5 shows various decorative interpretations of the curve Euler (Cornu) spiral ($\alpha = -1$) shown in Fig. 2 (a). The curve is taken as a path along which various forms were distributed and grouped (using such composition tools as "rhythm" and "colour"). In these examples, the decorative character of the curve is its leading feature, which does not meet the requirement of unity in function and form (as in design).

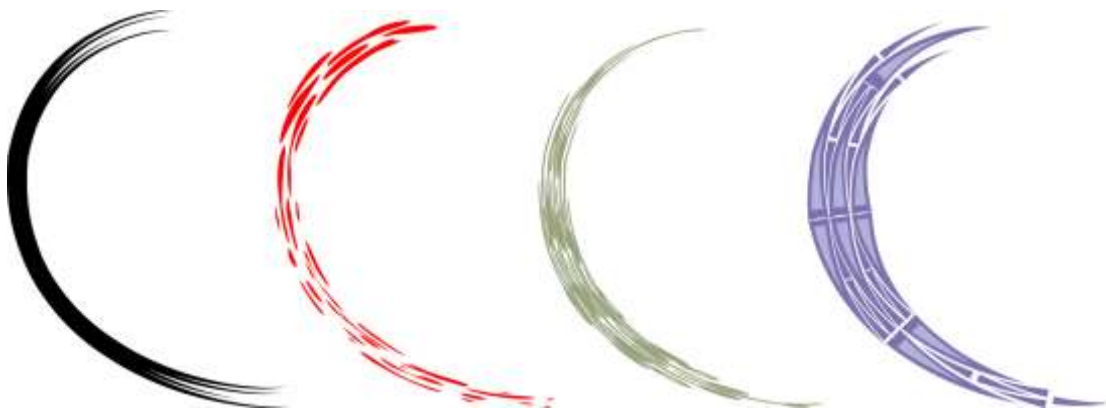

Fig. 5. Ornamental interpretation of Euler (Cornu) spiral ($\alpha = -1$) based on structuring and grouping of geometrical primitives along the spiral's curve.

Figures 6 and 7 evidence the existence of technical object form-generation, which reveals properties of analysed curves.

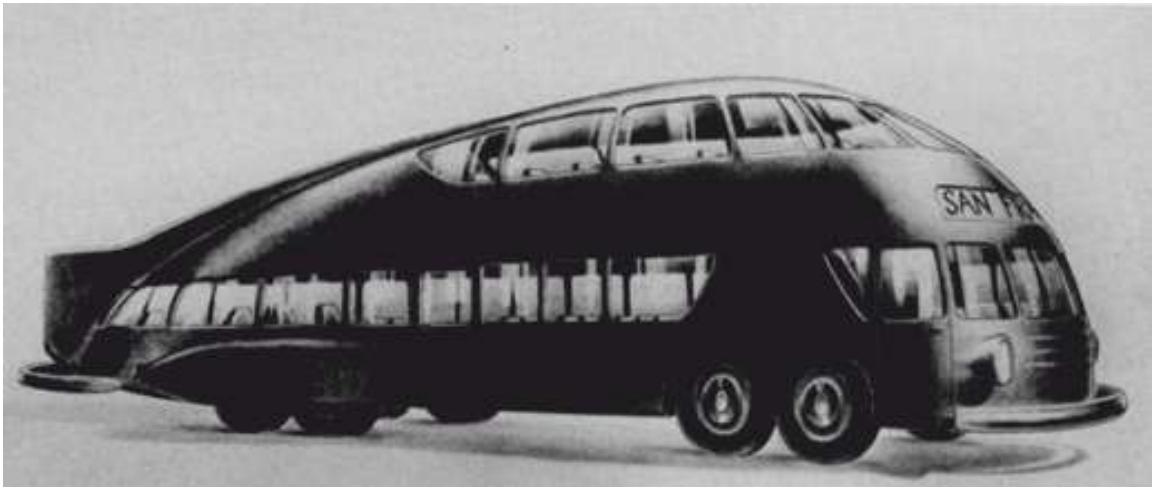
Fig. 6. Norman Bel Geddes. Rapid intercity bus.

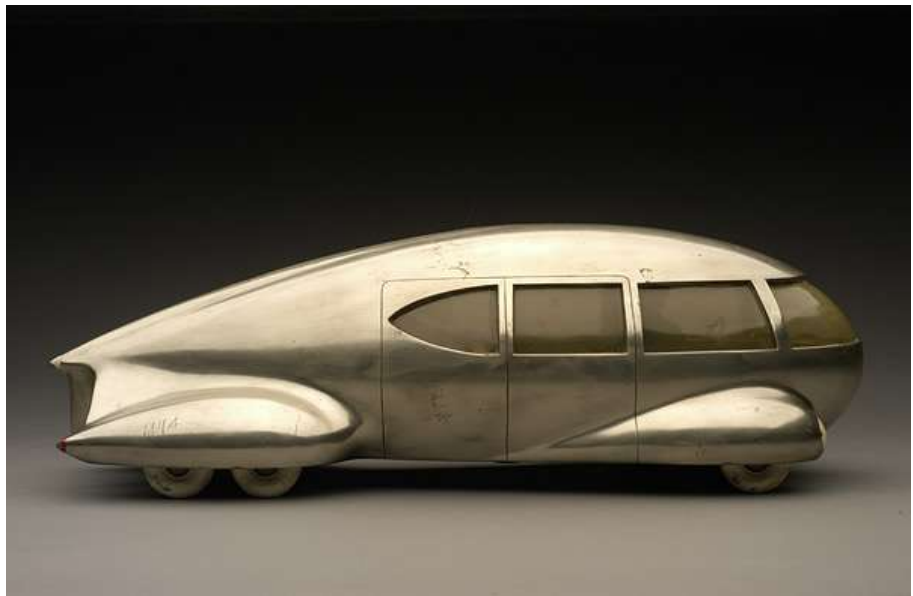
Fig. 7. Norman Bel Geddes. Model of streamlined "car of the future", 1933.

### 7. Conclusion

Plane curves with monotone function of curvature are widely applied in computer geometric design and computer graphics. The field of their application is quite broad, ranging from font design to simulation of aircrafts' and ballistic rockets' surfaces.

The performed aesthetic analysis and evaluation of plastic properties of presented curves from the viewpoint of technical aesthetics vividly show the practicability of using tools for mathematic modelling of geometrical forms. Complex industrial products utilizing functions of knowledge-intensive design involving mathematical methods should embody unity of usefulness and beauty as a condition of harmonization of natural and artificial matters. This principle may not be implemented without formation of geometrical structures and their analysis from the viewpoint of aesthetic practicability, which is dialectically integrated in any matter or phenomenon.

Aesthetic analysis does not cover all aspects of the problem in question and offers very promising prospects for further development; for example, this method can be used for curve estimation in a 3D Cartesian coordinate system.

Thus, our analysis demonstrates the ability to implement the classical concept of science and art combined to produce an integrated approach to geometrical shaping, realizing its full potential both for design and engineering purposes.

In our opinion, development of algorithms to generate and control the curvature for class A Bézier curves, as basic functions containing generalized Bernstein polynomials, i.e., Stancu- [20], Lupas- [21] or Vidensky-type linear operators [22], seems to offer great opportunities. We will address these issues in our further work.